\documentclass[twocolumn,prb,superscriptaddress,a4paper,floatfix,showkeys]{revtex4}
\usepackage{graphicx}
\usepackage{amsmath}

\begin{document}

\title{Quantum annealing: An introduction and new developments}
\author{Masayuki Ohzeki}
\affiliation{Department of Systems Science, Kyoto University, Yoshida-Honmachi, Sakyo-ku,
Kyoto, 606-8501, Japan}
\author{Hidetoshi Nishimori }
\affiliation{Department of Physics, Tokyo Institute of Technology, Oh-okayama, Meguro-ku,
Tokyo 152-8551, Japan}
\date{\today }

\begin{abstract}
Quantum annealing is a generic algorithm using quantum-mechanical
fluctuations to search for the solution of an optimization problem. The
present paper first reviews the fundamentals of quantum annealing and then
reports on preliminary results for an alternative method. The review part
includes the relationship of quantum annealing with classical simulated
annealing. We next propose a novel quantum algorithm which might be
available for hard optimization problems by using a classical-quantum
mapping as well as the Jarzynski equality introduced in nonequilibrium
statistical physics.\newline
\end{abstract}

\maketitle

\section{Introduction}

Recent remarkable advances in experimental techniques have enabled us to
fabricate dedicated nanostructures, in which prominent quantum effects are
observed and/or controlled. Such technological achievements open up the
possibilities of realization of advanced methods of quantum computation. In
this article, we review and present new developments in quantum annealing
(QA) \cite{QA1,QA2,QA3,QA4,QA5,QA6,QA} as an important subfield of quantum
computation.

Quantum annealing is a generic algorithm intended for solving optimization
problems by use of quantum tunneling. An optimization problem is formulated
as a task to minimize a real single-valued function, called a cost function,
of multivariables. We will in particular be interested in combinatorial
optimization problems, in which variables assume discrete values \cite%
{OP1,OP2}. We will use the simple term of ``an optimization problem" to mean
a combinatorial optimization problem.

Optimization problems are classified roughly into two types, easy and hard
ones. They are distinguished by the computational cost to solve optimization
problems. Easy problems are those which can be solved by the best algorithms
in time polynomial in the problem size (polynomial complexity). For hard
problems, in contrast, the best known algorithms cost exponentially long
time as a function of the system size (exponential complexity). For these
latter problems, it is virtually impossible to find the optimal solution in
a reasonable time, if the problem size exceeds a moderate value. Most of the
interesting optimization problems belong to the latter hard class. It is
thus important practically to devise algorithms which give approximate but
accurate solutions efficiently.

One of the generic algorithms proposed as part of such efforts is QA. This
is a method to use quantum dynamics and quantum tunneling effects to let the
system explore the phase space toward the solution of an optimization
problem. Quantum adiabatic evolution (or quantum adiabatic computation)
shares essentially the same idea \cite{Farhi}.

A related algorithm of simulated annealing (SA) is widely used in practice
to solve optimization problems approximately in a reasonable time \cite%
{SA1,SA2}. In SA, the cost function to be minimized is identified with the
energy of a classical statistical-mechanical system. The system is then
given a temperature, as an artificial control parameter. By decreasing the
parameter slowly from a high value to zero, one hopes to drive the system to
the state with the lowest value of the energy. The idea is that the system
is expected to stay close to thermal equilibrium during the protocol in SA.
If the rate of decrease in temperature is sufficiently slow, the system will
be led in the end to the zero-temperature equilibrium state, the
lowest-energy state. SA is thus regarded as an algorithm to make use of
(non-)equilibrium statistical mechanics for an efficient exploration of the
phase space.

Let us turn our attention to QA, which is the main topic of this article. In
SA, we make use of thermal, classical, fluctuations to introduce a
stochastic search for the desired lowest-energy state by allowing the system
to hop from state to state over intermediate energy barriers. In QA, by
contrast, we introduce non-commutative operators as artificial degrees of
freedom of quantum nature. They induce quantum fluctuations. We first set
the strength of quantum fluctuations to a very large value to search for the
global structure of the phase space, similarly to the high-temperature
situation in SA. Then the strength is gradually decreased to finally vanish
to recover the original system hopefully in the lowest-energy state. Quantum
tunneling between different classical states replaces thermal hopping in SA.

An observant reader should have noticed a number of similarities and
differences between SA and QA. In both methods, we have to control the
relevant parameters slowly and carefully to tune the strengths of thermal or
quantum fluctuations as desired. The idea behind QA is to keep the system
close to the instantaneous ground state of a quantum system. This aspect is
analogous to the protocol of SA, in which one tries to keep the system state
in quasi-equilibrium. It is indeed possible to render this analogy into a
more precise formulation, from which several important developments have
been achieved. An example of such cases will be explained later.

Let us move our attention to some of the developments in nonequilibrium
statistical physics. Several exact relations between nonequilibrium
processes and equilibrium states have been derived. The Jarzynski equality
(JE) is one of such remarkable results \cite{JE1,JE2}. This equality states
that we can reproduce useful information on equilibrium states from data
obtained by the repetition of non-equilibrium processes. We first prepare an
ensemble of states consisting of equilibrium distribution. We then drive
these states by tuning some parameters following a predetermined schedule.
In general the system during this process cannot be kept in equilibrium.
However, the average with weights given by the exponentiated value of the
work done during the nonequilibrium process coincides with the thermal
average of the exponential of free energy difference between the final and
initial states in equilibrium. The state is not necessarily kept in
equilibrium also during the protocol of SA as mentioned above. However, JE
tells us the possibility to obtain information of equilibrium state from
nonequilibrium processes. Indeed, the application of JE to SA has been
studied by several researchers \cite{NJ,Iba,Pop1,ON1,ON2}. This application
is known to successfully extract information on the equilibrium state faster
than the ordinary procedure of SA with the aid of the property of JE. We
show a quantum version of the algorithm with JE in this article.

The present paper consists of five sections. In \S 2, we review the
foundation of QA. In the following section, we introduce the
classical-quantum mapping. Here we understand the property of SA from the
point of view of quantum mechanics. Then, we construct an algorithm of QA
with JE in \S 4. The last section presents a summary.

\section{Quantum annealing}

Suppose that the optimization problem we wish to solve has been formulated
as the minimization of a cost function, which we regard as a Hamiltonian $%
H_{0}$ of a classical many-body system. As a simple example, let us consider
the problem of database search to find a specific item among $N$ items. If
the specific item to be found is denoted by a bracket $|\Psi _{0}\rangle $,
the cost function to be minimized is simply written as 
\begin{equation}
H_{0}=I-|\Psi _{0}\rangle \langle \Psi _{0}|,  \label{H0_Grover}
\end{equation}%
where $I$ is the $N$-dimensional identity matrix.

\subsection{Algorithm of QA}

In QA, we use quantum fluctuations to search the ground state of $H_{0}$. To
this end we introduce an operator $H_{1}$, which does not commute with $%
H_{0} $. The whole quantum system is then expressed as 
\begin{equation}
H(t)=\frac{t}{\tau }H_{0}+\left( 1-\frac{t}{\tau }\right) H_{1},  \label{Glo}
\end{equation}%
where $\tau$ is the computation time and is an important parameter in the
following discussions. The initial Hamiltonian is set as $H(0)=H_{1}$ and
the final system is described by $H(\tau)=H_{0}$. It is crucial to prepare
an appropriate Hamiltonian $H_{1}$, whose ground state is trivially found.
For instance, we prepare $H_{1}$ with the following form, 
\begin{equation}
H_{1}=I-|\Phi _{0}\rangle \langle \Phi _{0}|,
\end{equation}%
where 
\begin{equation}
|\Phi _{0}\rangle =\frac{1}{\sqrt{N}}\sum_{j=1}|j\rangle .
\label{Initial_state_Grover}
\end{equation}%
Its ground state is trivially given by a linear combination of all states
with a uniform weight.

The optimization protocol by QA is to start the algorithm from the
preparation of the ground state of $H_{1}$ as the initial state. The whole
system is then driven by the time-dependent Hamiltonian $H(t)$ following the
Schr\"odinger equation. If we choose a very large value of $\tau $, the
system evolves slowly. Such a slow time evolution would let the system
closely follow the instantaneous ground state of the total Hamiltonian $H(t)$
if the condition set by the adiabatic theorem is satisfied.

\subsection{Adiabatic evolution}

Let us formulate the above statement more precisely. For a sufficiently
large $\tau$, the adiabatic theorem guarantees that the state at time $t$, $%
|\psi (t)\rangle$, is very close to the instantaneous ground state $%
|0(t)\rangle $, $|\langle 0(t)|\psi(t)\rangle |\approx 1$, if the following
condition is satisfied, in consideration that the protocol of QA starts from
the initial ground state. We assume that the instantaneous ground state $%
|0(t)\rangle $ is non-degenerate. If we write the instantaneous first
excited state and its energy as $|1(t)\rangle $ and $E_{1}(t)$,
respectively, the quantum adiabatic theorem states that condition for $%
|\langle 0(t)|\psi (t)\rangle|^2=1-\epsilon ^{2}$ ($\epsilon \ll 1$) to hold
is given by 
\begin{equation}
\frac{\mathrm{max}\left|\langle 1(t)|\displaystyle\frac{dH(t)}{dt}%
|0(t)\rangle\right| }{\mathrm{min}\Delta (t)^{2}}=\epsilon ,
\label{adiabatic_condition}
\end{equation}%
where max and min are evaluated between $t=0$ and $t=\tau $ and $\Delta (t)$
is the energy gap, $\Delta (t)=E_{1}(t)-E_{0}(t)$ with $E_{0}(t)$ being the
instantaneous ground-state energy. For our case of Eq. (\ref{Glo}), the
numerator of Eq. (\ref{adiabatic_condition}) with the time derivative of $%
H(t)$ is proportional to $1/\tau $. Therefore the adiabatic condition of QA
is written as 
\begin{equation}
\tau \propto \frac{1}{\epsilon ~\mathrm{min}\Delta (t)^{2}}.
\label{computation_time}
\end{equation}%
If we choose $\tau $ large enough to satisfy this condition, we can obtain
the ground state of $H_{0}$ at $t=\tau$ with a high probability. A serious
problem arises if the energy gap $\Delta (t)$ collapses as the system size
increases because then the computation time $\tau$ increases according to
Eq. (\ref{computation_time}). Also the requirement of a small error
probability $\epsilon$ pushes $\tau$ to a larger value.

We can evaluate the computation time of the database search problem by QA
using Eqs. (\ref{H0_Grover})-(\ref{Initial_state_Grover}) \cite{GloG}. The
energy gap is given by 
\begin{equation}
\Delta (t)=\sqrt{1-\frac{1}{4}\left( 1-\frac{1}{N}\right) \frac{t}{\tau }%
\left( 1-\frac{t}{\tau }\right) }.
\end{equation}
Its minimum is $\mathrm{min}\Delta (\tau /2)=1/\sqrt{N}$. Therefore the
computation time is proportional to $N$. This is essentially the same as the
simple serial search. However, if we control the system evolution slowly
around $t=\tau/2$, where the gap assumes its minimum, than at other time,
then the transition to excited states is likely to be suppressed and the
probability of success may increase. More explicitly, the time dependence of
the Hamiltonian is chosen to involve a monotonically increasing function $%
f(t)$, instead of the simple $t/\tau$ and $1-t/\tau$ in Eq. (\ref{Glo}), as 
\begin{equation}
H(t)=f(t)H_{0}+\left\{ 1-f(t)\right\} H_{1}.  \label{QA}
\end{equation}
Then we demand that the adiabatic condition of Eq. (\ref{adiabatic_condition}%
), without the `max' and `min' operations, is satisfied at each time $t$.
The result is a differential equation for $f(t)$, by solving which we
achieve the optimal control of the dynamics. It actually turns out that the
total time evolution obtained by this method is proportional to $\sqrt{N}$.
This is a significant improvement over the above-mentioned simple result $%
\tau\propto N$. This is equivalent to the computation time of the Grover
algorithm \cite{Grover}. This is an explicit, remarkable example in which
quantum effects accelerate the convergence toward the optimal solution. It
has been established in many other cases by numerical and analytical methods
that QA achieves superior performance compared with SA \cite{CT}. Also known
is a convergence theorem of QA \cite{CT}: It is guaranteed that the system
converges to the optimal solution after an infinite time of evolution if an
appropriate control of a system parameter is chosen. This is to be compared
with a convergence theorem for SA \cite{GG}, where the time control of the
system parameter (temperature) is constrained to be slower than the
corresponding one in QA.

\subsection{Errors in QA}

Although the convergence of QA to the optimal solution is guaranteed in the
infinite-time limit as mentioned above, we have to carry out the computation
in a finite time $\tau$. It is therefore important to estimate how the
error, the probability not to reach the optimal state, depends upon $\tau$.
The adiabatic theorem provides an answer to this question: The error after
an adiabatic evolution of $\tau $ is generically proportional to $\tau ^{-2} 
$ in the limit of large $\tau $ as long as the system size $N$ is kept
finite \cite{SO}. However, if the system size grows, the energy gap $\Delta
(t)$ becomes small and eventually collapses to zero as $N\to\infty$ when a
quantum phase transition takes place. Thus the study of quantum phase
transitions is an important part of the analyses of QA.

For instance, how the error in QA depends on the complexity of the
optimization problems is interesting in order to understand the performance
of QA. The errors in the final state after the process of QA can be regarded
as imperfections or defects in the state caused by the evolution across a
phase transition. The dynamics across a phase transition is well understood
via the Kibble-Zurek mechanism \cite{KZ1,KZ2}. This aspect of error
evaluation has been attracting increasing attention \cite%
{KZ3,KZ4,KZ5,KZ6,KZ7}.

\subsection{Hard optimization problems}

A hard optimization problem is often represented in QA in terms of a
Hamiltonian $H(t)$ with a phase transition accompanying an exponentially
small energy gap $\propto \exp (-\alpha N)~(\alpha >0)$. One of the examples
is found in the random energy model \cite{REM}. Consider the Hamiltonian 
\begin{equation}
H_{0}=-\sum_{\{i_{p}\}}J_{i_{1},i_{2},\cdots ,i_{p}}\sigma
_{i_{1}}^{z}\sigma _{i_{2}}^{z}\cdots \sigma _{i_{p}}^{z},  \label{REM}
\end{equation}%
where the summation is taken over all combinations of $p$ sites among $N$ of
them, and $\sigma_i^z$ is the $z$ component of a Pauli spin operator at site 
$i$. The interaction $J_{i_{1},i_{2},\cdots ,i_{p}}$ follows the Gaussian
distribution with a vanishing mean and variance $p!/(2N^{p-1})$, which
guarantees a sensible thermodynamic limit for any $p$ including the limit $%
p\to\infty$. The random energy model is defined as the limit $p\to\infty$ of
the Hamiltonian (\ref{REM}). This limit significantly simplifies the
problem. In particular the energy spectrum turns out to obey the Gaussian
distribution with vanishing mean and variance $\sqrt{N/2}$. Any energy level
takes a value drawn from this distribution independently of other levels.

Let us consider to find the ground state of the random energy model by QA.
The transverse-field operator is often used as quantum fluctuations for
searching the ground state in spin glasses including the random energy
model, 
\begin{equation}
H_{1}=-\sum_{i}\sigma _{i}^{x}.
\end{equation}
Quantum annealing of the random energy model has the ground state of $H_{1}$
as the initial state, which is trivially given as $|\Psi(0)\rangle
=\sum_{\{\sigma\}}|\sigma \rangle /{\sqrt{2}^{N}}$, where $\sigma_i$ is an
eigenvalue of $\sigma_i^z$. By a perturbation theory, we can evaluate the
asymptotic behavior of the minimum energy gap of the random energy model
with a transverse field as $\Delta \propto\exp(-\alpha N)$ for a
sufficiently large system size $N$ \cite{FT}. Therefore the adiabatic
theorem states that QA, as long as one demands the system to stay
arbitrarily close to the instantaneous ground state, does not work in a time
polynomial in system size.

We introduce another instance from one of the famous combinatorial
optimization problems, the exact cover. This is a special version of
constraint satisfaction problems. Let us consider $N$ spins and $M$
``clauses", each one of which involves three Ising spins chosen at random.
The energy of a clause is zero if one spin is $-1$ and the other two are $1$%
. Otherwise, the energy is defined as 1. This assignment of the energy is
realized by an Ising model as 
\begin{eqnarray}
H_{0} &=&\frac{1}{8}\sum_{m=1}^{M}\left( \sigma _{m_{1}}^{z}\sigma
_{m_{2}}^{z}\sigma _{m_{3}}^{z}-\sigma _{m_{1}}^{z}-\sigma
_{m_{2}}^{z}-\sigma _{m_{3}}^{z}\right.  \notag \\
&&\quad \left. +\sigma _{m_{1}}^{z}\sigma _{m_{2}}^{z}+\sigma
_{m_{2}}^{z}\sigma _{m_{3}}^{z}+\sigma _{m_{3}}^{z}\sigma
_{m_{1}}^{z}+5\right) .  \notag \\
&&
\end{eqnarray}%
Numerical simulations have shown that the minimum energy gap during QA of
the exact cover has the asymptotic form proportional to $\exp (-\alpha N)$ 
\cite{FT2,FT3}. Thus we again encounter a similar problem to the random
energy model.

The behavior of the energy gap as $\exp(-\alpha N)$ is characteristic of
first-order phase transitions in quantum systems. Therefore the problem
concerning the bottleneck of QA appears even in the simple ferromagnetic
model, namely the Ising model with three-body interaction \cite{FT4}. This
model also exhibits a first-order phase transition with an exponentially
small gap. This fact may be taken as an indication that QA would fail to
find a trivial ground state in a reasonable time. However we have some room
in the choice of quantum fluctuations expressed by $H_{1}$. It remains to
see if an appropriate choice of the quantum term $H_{1}$ can help us avoid
these difficulties in the framework of QA under the constraint of an
adiabatic evolution.

In the rest of this article, we provide a different approach to overcome the
difficulty of the energy gap closure based on a different idea.

\section{Classical-quantum mapping}

The analogy between QA and SA can be pursued in a systematic manner by the
exploitation of an ingenious mapping between a classical thermal state and a
corresponding quantum state \cite{QC}. This mapping is different from the
conventional quantum-classical mapping in terms of the Suzuki-Trotter
formula or the path integral formulation. In particular, the same spatial
dimensionality is shared by classical and quantum systems according to the
present mapping.

\subsection{Mapping of a classical system to a quantum system}

In order to formulate the classical-quantum mapping, we consider a classical
Ising spin system with a configuration $\sigma $. This system is assumed to
obey a stochastic dynamics described by the master equation 
\begin{equation}
\frac{d}{dt}P(\sigma ;t)=\sum_{\sigma ^{\prime }}M(\sigma |\sigma ^{\prime
};t)P(\sigma ^{\prime };t),
\end{equation}%
where $M(\sigma ^{\prime }|\sigma ;t)$ expresses the transition matrix
satisfying the detailed balance condition 
\begin{equation}
M(\sigma |\sigma ^{\prime };t)\tilde{P}(\sigma ^{\prime })=M(\sigma ^{\prime
}|\sigma ;t)\tilde{P}(\sigma ).  \label{db}
\end{equation}%
The instantaneous equilibrium distribution $e^{-\beta(t) H_{0}(\sigma)}/Z(t)$
is denoted as $\tilde{P}(\sigma ;t)$. Since we are considering a dynamic
control of the temperature, $T(t)$, the time variable $t$ has been written
explicitly in the arguments of the inverse temperature and the partition
function.

The (instantaneous) equilibrium state of the above stochastic dynamics is
now shown to be expressed as the ground state of a quantum Hamiltonian. A
general form of such a quantum Hamiltonian is given as 
\begin{equation}
H_q(\sigma^{\prime }|\sigma;t) = \delta_{\sigma^{\prime },\sigma} - \mathrm{e%
}^{\beta(t)H_0(\sigma^{\prime })/2}M(\sigma^{\prime }|\sigma;t) \mathrm{e}%
^{-\beta(t)H_0(\sigma)/2}.  \label{Hq}
\end{equation}
This Hamiltonian has the following state as its ground state, 
\begin{equation}
|\Psi_{\mathrm{eq}}(t)\rangle =\frac{1}{\sqrt{Z(t)}}\sum_{\sigma }\mathrm{e}%
^{-\frac{\beta (t)}{2}H_{0}(\sigma)}|\sigma \rangle .  \label{Ground}
\end{equation}
It is clear that the quantum expectation value of a physical quantity $%
A(\sigma)$ by $|\Psi_{\mathrm{eq}}(t)\rangle$ coincides with the thermal
expectation value of the same quantity. The ground state energy for this
ground state is $0$, which is shown by the detailed-balance condition, 
\begin{eqnarray}
& & \left(\delta_{\sigma^{\prime },\sigma} - \mathrm{e}^{\frac{\beta(t)}{2}%
H_0(\sigma^{\prime })}M(\sigma^{\prime }|\sigma;t) \mathrm{e}^{-\frac{%
\beta(t)}{2}H_0(\sigma)}\right)|\Psi_{\mathrm{eq}}(t)\rangle  \notag \\
& & \propto\sum_{\sigma}\left(\mathrm{e}^{-\frac{\beta(t)}{2}%
H_0(\sigma^{\prime})} - \mathrm{e}^{\frac{-\beta(t)}{2}H_0(\sigma^{\prime
})}M(\sigma|\sigma^{\prime};t)\right)|\sigma\rangle  \notag \\
& & =0,
\end{eqnarray}
where we used the detailed balance condition and the requirement for
conservation of probability $\sum_{\sigma}M(\sigma|\sigma^{\prime
};t)|\sigma\rangle = |\sigma^{\prime}\rangle$. The excited states have
positive-definite eigenvalues, which can be confirmed by the application of
the Perron-Frobenius theorem. By using the above special quantum system, we
can treat a quasi-equilibrium stochastic process in SA as an adiabatic
quantum-mechanical dynamics in QA. The stochastic dynamics can be mapped
into the quantum dynamics as above introduced even for the system with
continuous variables by focusing the relationship between the Fokker-Planck
equation and the Shr\"odinger equation \cite{Orland}.

\subsection{Similarity of SA and QA}

Let us consider an optimization problem that can be expressed as a
Hamiltonian with local interactions 
\begin{equation}
H_{0}=\sum_{j}H_{j},  \label{H0_classical_j}
\end{equation}%
where $H_{j}$ involves $\sigma _{j}^{z}$ and a finite number of $\sigma
_{k}^{z}~(k\neq j)$. A simple instance is a system with nearest-neighbor
interactions, 
\begin{equation}
H_{j}=-\sum_{k\in j}J_{jk}\sigma _{j}^{z}\sigma _{k}^{z},
\end{equation}%
where $k$ is a site neighboring to $j$. The following explicit choice of the
quantum Hamiltonian as an example of Eq. (\ref{Hq}) facilitates our
analysis, 
\begin{equation}
H_{q}^{\mathrm{SG}}(t)=-\chi (t)\sum_{j}\left( \sigma _{j}^{x}-\mathrm{e}^{%
\frac{\beta (t)}{2}H_{j}}\right) ,  \label{Hq_SG}
\end{equation}%
where $\chi (t)=e^{-\beta (t)p}$ with $p=\mathrm{max}_{i}|H_{i}|$, which is
proportional to the energy and is of $\mathcal{O}(N^{0})$ because of
finiteness of the interaction range.

The quantum system (\ref{Hq_SG}) has the following interesting property. For
very high temperatures, $\beta\to 0$, this quantum Hamiltonian reduces 
\begin{equation}
H^{\mathrm{SG}}_{q}(t)=-\sum_{j}\left( \sigma _{j}^{x}- 1\right).
\label{Hq_SG1}
\end{equation}
The ground state of Eq. (\ref{Hq_SG1}) is the uniform linear combination of
all possible states in the basis to diagonalize $\{\sigma_i^z\}$, which
means that all states appear with an equal probability. Such a situation is
realized also in the high-temperature limit of the original classical system
represented by Eq. (\ref{H0_classical_j}). Thus the quantum ground state in
the limit $\beta\to 0$ appropriately describes the classical disordered
state. Analogously, in the limit $\beta\to\infty$, the quantum Hamiltonian
becomes 
\begin{equation}
H^{\mathrm{SG}}_{q}(t)\sim\chi(t)\sum_{j}\mathrm{e}^{\frac{\beta(t)}{2} H_j},
\end{equation}
whose ground state is the ground state of the classical system (\ref%
{H0_classical_j}) because each $H_j$ takes its lowest value. It has
therefore been confirmed that the limit $\beta(t)\to\infty$ of the quantum
system correctly corresponds to the low-temperature limit of the classical
system.

Surprisingly, the adiabatic condition applied to the quantum system (\ref%
{Hq_SG}) can be shown to lead to the condition of convergence of SA \cite{QC}%
. If we assume a monotonic decrease of the temperature $T(t)=1/\beta (t)$,
the adiabatic condition, Eq. (\ref{adiabatic_condition}) without max and
min, applied to the quantum Hamiltonian (\ref{Hq_SG}) yields the following
time dependence of $\beta (t)$ in the limit of large $N$, 
\begin{equation}
\beta (t)=\frac{\log (c t+1)}{pN}.  \label{SA_schedule}
\end{equation}%
The coefficient $c$ is exponentially small in $N$. The above result
reproduces the Geman-Geman condition for convergence of SA \cite{GG}. These
authors used the theory of the classical time-dependent Markov chain
representing nonequilibrium stochastic processes. The classical system under
such a time evolution may not seem to stay close to equilibrium since the
temperature changes continually. However, the result mentioned above implies
that the quasi-equilibrium condition is actually satisfied during the
process following Eq. (\ref{SA_schedule}) because the adiabatic condition
means that the quantum system is always very close to the instantaneous
ground state, which represents the equilibrium state of the original
classical system.

The classical-quantum mapping explained in this section will be shown to be
an important ingredient of the development in a later section.

\section{Jarzynski equality}

We now develop a new algorithm for optimization problems using the
classical-quantum mapping and the Jarzynski equality. This latter equality
relates quantities at two different thermal equilibrium states with those of
non-equilibrium processes connecting these two states. It can also be
regarded as a generalization of the well-known inequality of thermodynamics, 
$\Delta F \le\langle W\rangle$. Here the brackets $\langle \cdots \rangle$
are for the average taken over non-equilibrium processes between the initial
and final states, both of which are specified only macroscopically and thus
there can be a number of microscopic realizations.

\subsection{Formulation of the Jarzynski equality}

The Jarzynski equality is written as 
\begin{equation}
\left\langle \mathrm{e}^{-\beta W} \right\rangle = \frac{Z_{\tau}}{Z_{0}}.
\end{equation}
Here the partition functions for the initial and final Hamiltonians are
expressed as $Z_{0}$ and $Z_{\tau}$, respectively. One of the important
features is that JE holds independently of the pre-determined schedule of
the nonequilibrium process. Another celebrated benefit is that JE reproduces
the second law of thermodynamics given as $\Delta F\le \langle W\rangle$ by
using the Jensen inequality, $\langle e^{-\beta W} \rangle \ge e^{-\beta
\langle W\rangle}$. Notice that we have to take all fluctuations into
account in formation of the expectation value in the right-hand side of JE,
if one uses the calculation of the free energy difference in practical use.
Finite errors on the ratio of the partition functions are attributed to the
rare event on nonequilibrium processes \cite{Ref1,Ref2,Ref3}.

We show the proof of JE for a classical system in contact with a heat bath 
\cite{JE2}. Let us consider a thermal nonequilibrium process in a
finite-time schedule $0\leq t\leq \tau $. Thermal fluctuations are simulated
by the master equation. We employ discrete time expressions and write $%
t_{k+1}-t_{k}=\delta t$, $t_{0}=0$ and $t_{n}=\tau $. The probability that
the system is in a state $\sigma_k$ at time $t_k$ will be denoted as $%
P(\sigma _{k};t_{k})$. Notice that $\sigma_k$ is not a state of a single
spin at site $k$ but is a collection of spin states at time $t_k$. The
transition probability per unit time is written as $M(\sigma _{k+1}|\sigma
_{k};t_{k})$. In the original formulation of JE, the work is defined as the
energy difference due solely to the change of the Hamiltonian. We can
construct JE in the case of changing the inverse temperature, which is
useful in the following application, by defining the work as $-\beta
W(\sigma _{k};t_{k}) = - (\beta(t_{k+1})- \beta(t_k)) E(\sigma_{k})$, where $%
E(\sigma)$ is the value of the cost function (classical Hamiltonian $%
H_0(\sigma)$) with the spin configuration $\sigma$. The left-hand side of JE
is then expressed as 
\begin{eqnarray}
\left\langle \mathrm{e}^{-\beta W}\right\rangle &=&\sum_{\{\sigma
_{k}\}}\prod_{k=0}^{n-1}\left\{ \mathrm{e}^{-\beta W(\sigma _{k+1};t_{k})}%
\mathrm{e}^{\delta tM(\sigma _{k+1}|\sigma _{k};t_{k})}\right\}  \notag \\
&&\quad \times \tilde{P}(\sigma _{0};t_{0}).  \label{CJE}
\end{eqnarray}%
This means that the system evolves from state $\sigma_k$ to state $%
\sigma_{k+1}$ during the interval $\delta t$ and then we measure the work $%
W(\sigma_{k+1};t_{k})$. This process is repeated from $k=0$ to $n-1$. The
initial condition is set to the equilibrium distribution. If the transition
term $\exp (\delta tM(\sigma _{k+1}|\sigma_{k};t_{k}))$ is removed in this
equation, JE is trivially satisfied because the summation of $-\beta
W(\sigma_{k+1};t_k)$ over $k$ yields $-(\beta(t_n)-\beta(t_0))E(\sigma_0))$.
A non-trivial aspect of JE is that the insertion of the transition term does
not alter the conclusion.

Let us evaluate the first contribution $k=0$ of the product in the above
equation as 
\begin{eqnarray}
&&\sum_{\sigma _{0}}\left\{ \mathrm{e}^{-\beta W(\sigma _{1};t_{0})}\mathrm{e%
}^{\delta tM_{0}(\sigma _{1}|\sigma _{0};t_{0})}\right\} \tilde{P}(\sigma
_{0};t_{0})  \notag \\
&&\quad =\frac{1}{Z_0}e^{-(\beta(t_1) -
\beta(t_0))E(\sigma_1)}e^{-\beta(t_0) E(\sigma_1)}  \notag \\
&&\quad =\tilde{P}(\sigma _{1};t_{1})\frac{Z_{1}}{Z_{0}}.
\end{eqnarray}%
Repeating the same manipulations, we obtain the quantity on the right-hand
side of Eq. (\ref{CJE}) as 
\begin{equation}
\sum_{\sigma _{n}}\tilde{P}(\sigma _{n};t_{n})\prod_{k=0}^{n-1}\frac{Z_{k+1}%
}{Z_{k}}=\frac{Z_{\tau }}{Z_{0}}.
\end{equation}

\subsection{Quantum Jarzynski Annealing}

Let us consider the application of the above formulation of JE to QA using
the quantum system (\ref{Hq}). Initially we prepare the trivial ground state
with a uniform linear combination as in ordinary QA. From the point of view
of the classical-quantum mapping, this initial state expresses the
high-temperature equilibrium state $|\Psi_{\mathrm{eq}} (t_0)\rangle \propto%
\mathrm{e}^{-\beta (t_{0})H_{0}(\sigma)/2}|\sigma \rangle $ with $\beta
(t_0)\ll 1$. Then we introduce the exponentiated work operator $%
W(\sigma_{k};t_k)=\exp (-(\beta (t_{k+1}) -\beta (t_{k}))H_{0}(\sigma_k)/2)$%
. If we apply this operation to the quantum wave function $|\Psi_{\mathrm{eq}%
} (t_0)\rangle$, the state is changed into a state corresponding to the
equilibrium distribution with the inverse temperature $\beta (t_{k+1})$.
Even if the quantum time-evolution operator $U(\sigma^{\prime
}|\sigma;t_{k+1})=\exp (-\mathrm{i}H_{q}(\sigma^{\prime
}|\sigma;t_{k+1})\delta t)$ is applied, this state does not change, since it
is the eigenstate of $H_{q}(\sigma^{\prime }|\sigma;t_{k+1})$. The obtained
state after the repetition of the above procedure is given as 
\begin{eqnarray}
|\Psi (t_{n})\rangle & \propto & \prod_{k=0}^{n-1}\left\{
W(\sigma_{k+1};t_{k})U_{k+1}(\sigma_{k+1}|\sigma_{k};t_{k}) \right\}  \notag
\\
& & \quad \times \left( \sum_{\sigma }\mathrm{e}^{-\frac{\beta(t_{0})}{2}%
H_{0}}|\sigma \rangle \right) .
\end{eqnarray}
This is essentially of the same form as Eq. (\ref{CJE}). Instead of $%
M(\sigma_{k+1}|\sigma_{k};t_{k})$, we use the time-evolution operator $%
U(\sigma_{k+1}|\sigma_{k};t_{k})$ here. After the system reaches the state $%
|\Psi (t_n)\rangle$, we measure the obtained state by the projection onto a
spin configuration $\sigma ^{\prime }$. The probability is then given by $%
|\langle \sigma ^{\prime }|\Psi (t_{n})\rangle |^{2}$. The ground state is
obtained by the probability proportional to $\exp(-\beta(\tau)H_{0})$, since 
$|\Psi (t_{n})\rangle \propto |\Psi_{\mathrm{eq}} (t_{n})\rangle$. If we
repeat the above procedure for $\beta (\tau)\gg 1$, we can efficiently
obtain the ground state of $H_{0}$. This is called the quantum Jarzynski
annealing (QJA) in this paper.

It may seem unnecessary to apply the time-evolution operator $U(\sigma
_{k+1}|\sigma _{k};t_{k})$ at the middle step between the operations of work
exponentiated operators $W(\sigma _{k+1};t_{k})$. In the formulation of JE,
we can use the identity matrix instead of $U(\sigma _{k+1}|\sigma _{k};t_{k})
$, but this is not perfectly controllable in quantum computation. The
quantum system always changes following the Shr\"{o}dinger dynamics. The
change by quantum fluctuations during the protocol is described by the time
evolution operator $U(\sigma _{k+1}|\sigma _{k};t_{k})$ at the middle step
between the operations of work exponentiated operators $W(\sigma
_{k+1};t_{k})$. However, it is useful to remember the nontrivial point of
JE. When transitions by quantum fluctuations between the exponentiated work
occurs, JE holds as in Eq. (\ref{CJE}).

We emphasize that the scheme of QJA does not rely on a quantum adiabatic
control. Aside from the number of the unitary gates to build the quantum
circuit of QJA, the computation time does not depend on the energy gap.
Therefore QJA would not suffer from the energy-gap closure differently from
ordinary QA. The result should be independent of the schedule of tuning the
parameter $\beta(t)$ in the above manipulations.

\subsection{Performance of QJA}

We expect that QJA gives the probability of the ground state following the
Gibbs-Boltzmann factor under any annealing schedule. In contrast, without
the multiplication of the exponentiated work, slow quantum sweep is
necessary for an efficient achievement to find the ground state according to
the ordinary QA. Let us consider two instances of the application of QJA.

\begin{figure}[tbp]
\begin{center}
\includegraphics[width=70mm]{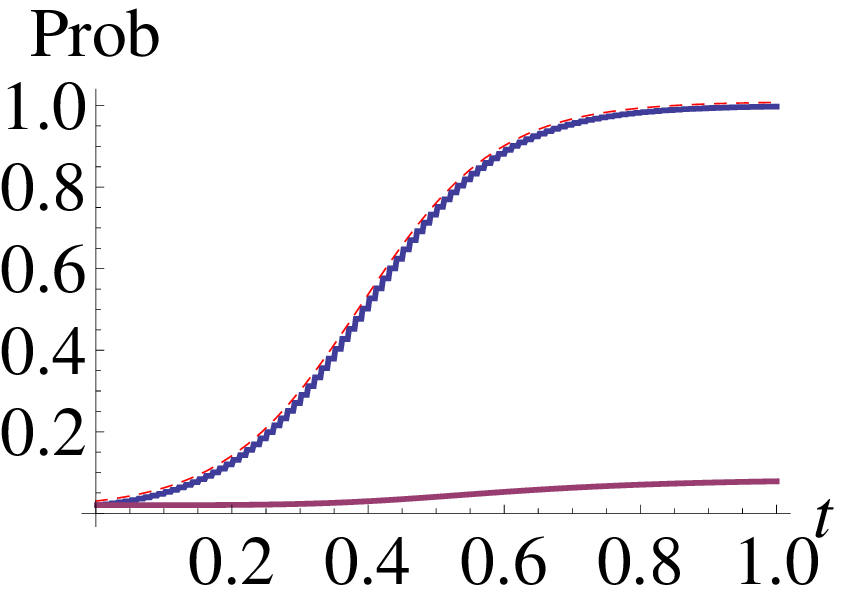} %
\includegraphics[width=70mm]{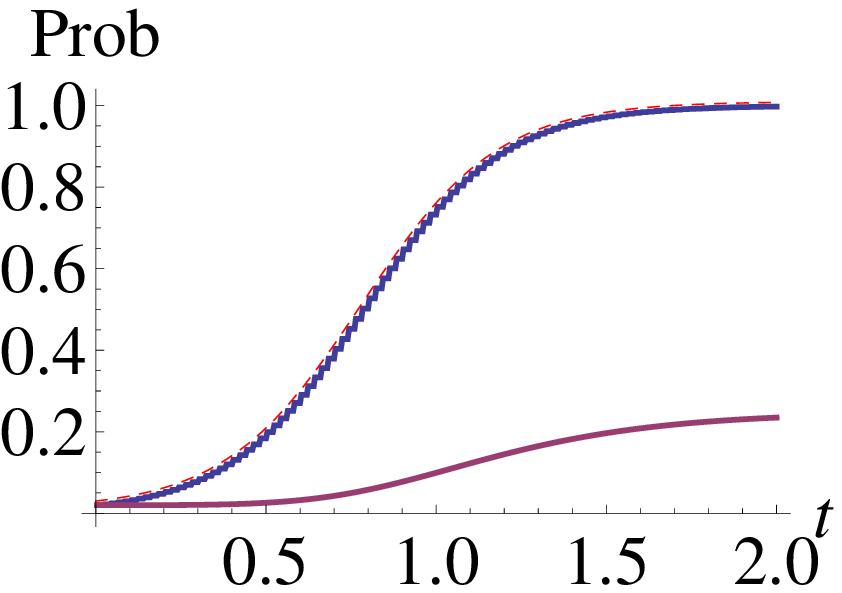} %
\includegraphics[width=70mm]{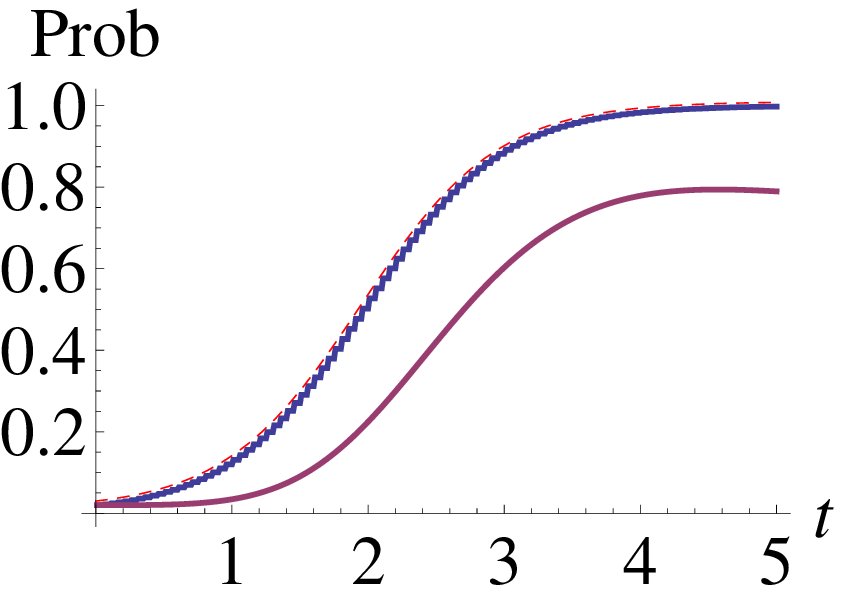}
\end{center}
\caption{{\protect\small Quantum Jarzynski annealing for the database search
problem from $N=50$ items ($N$ denotes the size of the Hilbert space). The
probabilities to find the desired state are plotted for $\protect\tau =1,2$,
and $5$ from top to bottom. The dashed curves denote the instantaneous
Gibbs-Boltzmann factor for reference. The upper solid curves (blue curves)
representing the results by QJA are fixed to these reference curves, while
the lower ones (red ones) express the time-dpendent results by the ordinary
QA.}}
\label{QJAG}
\end{figure}

The first example is the database search to find a specific item among $N=50$
items as explained in Sec. II. We reformulate this database search as a
single minimum potential among uniform values as $V_{0}=-1$, and $%
V_{i}=0~(i\neq 0)$. We construct the transition matrix following the Glauber
dynamics. We set up QA by the classical-quantum mapping and QJA by use the
same transition matrix for comparison. We increased $\beta (t)$ from $0$ to $%
10$ linearly as a function of $t$. Figure \ref{QJAG} describes the results
by QA and QJA for different computation time $\tau =1,2$ and $5$. 
The plots by QJA (upper curves) are indistinguishable from the reference
curve representing the instantaneous Gibbs-Boltzmann factor for the minimum
state $i=0$. In contrast, one finds that QA (lower curves) needs
sufficiently slow control of the quantum fluctuations to efficiently find
the minimum.

\begin{figure}[tbp]
\begin{center}
\includegraphics[width=70mm]{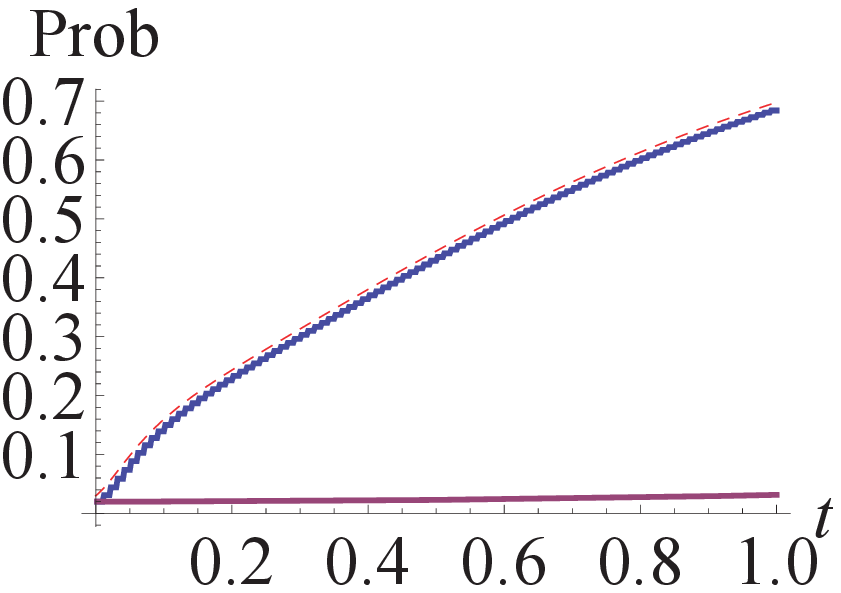} %
\includegraphics[width=70mm]{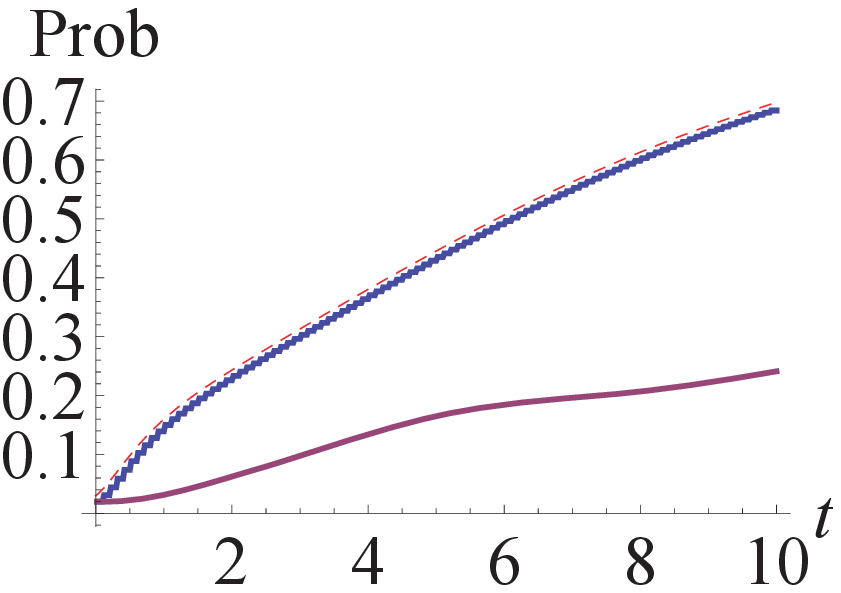} %
\includegraphics[width=70mm]{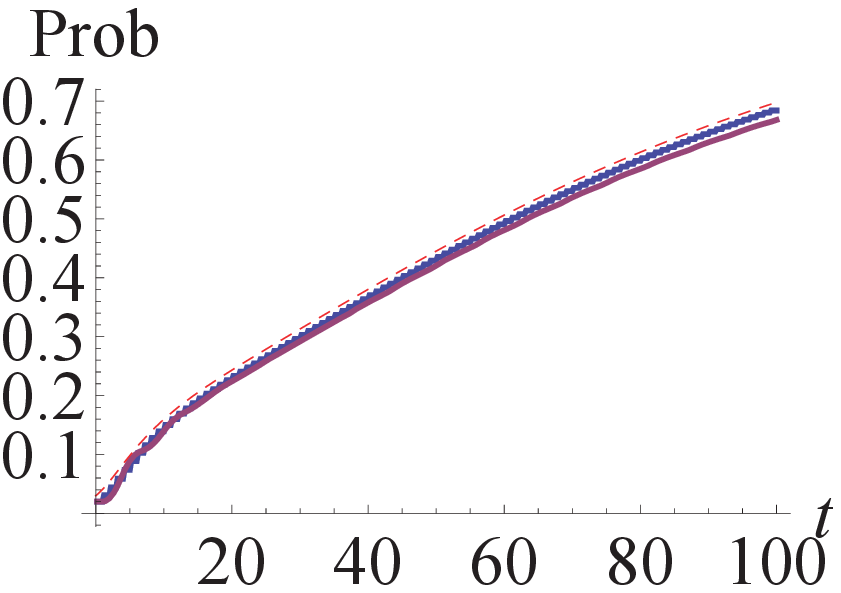}
\end{center}
\caption{{\protect\small Quantum Jarzynski annealing for the
random-potential problem for $N=50$ sites ($N$ denotes the size of the
Hilbert space). The probabilities to find the minimum-potential site are
plotted for $\protect\tau =1,10$, and $100$ from top to bottom. The same
symbols as in Fig. \protect\ref{QJAG}} are used.}
\label{QJARP}
\end{figure}

The second example is to search the minimum from a one-dimensional random
potential. We consider to find the minimum of the Hamiltonian $%
H_{0}=-\sum_{i=1}^{N}V_{i}|i\rangle \langle i|$, where $V_{i}$ denotes the
potential energy at site $i$. We choose $\{V_{i}\}$ randomly. We again
compare the case for $N=50$ sites by QJA and QA. By a linear schedule for
increasing the parameter $\beta(t) $ from $0$ to $100$, we carry out QA
without the exponentiated work operations and QJA. Figure \ref{QJARP} shows
the comparison between the probability for finding the ground state by QA
and QJA for different schedules $\tau =1,10$, and $100$. 
On finds that the plots by QJA (upper curves) are again indistinguishable
from the reference curve. QA (lower curves) needs sufficiently slow decrease
of quantum fluctuations to efficiently find the site with the minimum
potential.

QJA does not suffer from the energy-gap closure by increasing the system
size and gives the probability to obtain the ground state independently of a
predetermined schedule. Both of the above cases indeed show that the results
of QJA does not depend on the computational time $\tau$.

To realize QJA in an actual quantum computation, we need to implement the
exponentiated work operation $W(\sigma _{k};t_{k})$, since it looks like a
non-unitary operator. We can construct this operation as a unitary operation
by using an extra quantum system \cite{WQ,Ohzeki}. Instead of cut
\textquotedblleft time", we need another resource like a \textquotedblleft
memory" in quantum computation. Fortunately, the amount of needed memory
does not diverge exponentially by an increase of the system size $N$. Thus
the results by QJA shown here imply that we may overcome the difficulties in
hard optimization problems and solve them in a reasonable time. However, the
above instances do not represent hard optimization problems as in the
preceding section with an exponentially vanishing energy gap. In that sense,
our results are preliminary and future work should clarify the efficiency of
QJA for such hard problems.

\section{Summary}

In this paper we first reviewed QA, which is a generic algorithm to
approximately find the solution to optimization problems. The method is
based on the adiabatic evolution of quantum systems. A sufficient condition
to efficiently find the ground state by QA is given by the inverse square of
the energy gap. The same analysis in conjunction with the classical-quantum
mapping enables us to reproduce the convergence condition of SA, which is
another approximate solver exploiting thermal fluctuations. Recent
investigations disclosed the bottleneck of QA, the exponential closure of
the energy gap. According to the prediction by the adiabatic theorem, QA
does not work efficiently on problems with such a bottleneck. We thus have
to invent an alternative way to efficiently find the optimal solution with
the energy gap vanishing exponentially with the system size.

We next showed an application of JE to QA as one of the possible
improvements of QA, using the fact that we can map a quantum system close to
the instantaneous ground state to a thermal classical system in
quasi-equilibrium by the classical-quantum mapping. This protocol keeps the
quantum system to be the ground state expressing the instantaneous
equilibrium state. The cost for the realization of QJA in quantum
computation does not diverge exponentially as the system size increases,
which is the essentially different point from ordinary QA. This idea has
been shown to work in a few examples with a small Hilbert space. It is a
future problem to test this method for problems of much larger scales and
hard ones with the exponential closure of the energy gap.

\begin{acknowledgments}
This work was partially supported by CREST, JST, and scientific research on
the priority area by the Ministry of Education, Science, Sports and Culture
of Japan, ``Deepening and Expansion of Statistical Mechanical Informatics
(DEX-SMI)."
\end{acknowledgments}

\end{document}